\documentclass[twocolumn,aps,prl,showpacs,superscriptaddress]{revtex4}

\usepackage{graphicx}
\usepackage{amsfonts}
\usepackage{amsmath}
\usepackage{amssymb}
\usepackage{latexsym}
\usepackage{longtable}
\usepackage{ulem,color}

\newcommand{\ket} [1] {\vert #1 \rangle}

\begin{document}


\title{State-independent quantum contextuality with single photons}


\author{Elias Amselem}
 \affiliation{Department of Physics, Stockholm University, S-10691,
 Stockholm, Sweden}
\author{Magnus R{\aa}dmark}
 \affiliation{Department of Physics, Stockholm University, S-10691,
 Stockholm, Sweden}
\author{Mohamed Bourennane}
 \affiliation{Department of Physics, Stockholm University, S-10691,
 Stockholm, Sweden}
\author{Ad\'an Cabello}
 \affiliation{Departamento de F\'{\i}sica Aplicada II, Universidad de
 Sevilla, E-41012 Sevilla, Spain}


\date{\today}



\begin{abstract}
We present an experimental state-independent violation of an
inequality for noncontextual theories on single particles. We show
that 20 different single-photon states violate an inequality which
involves correlations between results of sequential compatible
measurements by at least 419 standard deviations. Our results show
that, for any physical system, even for a single system, and
independent of its state, there is a universal set of tests whose
results do not admit a noncontextual interpretation. This sheds new
light on the role of quantum mechanics in quantum information
processing.
\end{abstract}



\maketitle


The debate on whether quantum mechanics can be completed with hidden
variables started in 1935 with an ingenious example proposed by
Einstein, Podolsky, and Rosen \cite{EPR35} (EPR), suggesting that
quantum mechanics only gives an incomplete description of nature.
Schr\"odinger pointed out the fundamental role of quantum
entanglement in EPR's example and concluded that entanglement is
``{\it the} characteristic trait of quantum mechanics''
\cite{Schrodinger35}. For years, this has been a commonly accepted
paradigm, stimulated by the impact of the applications of
entanglement in quantum communication \cite{BW92, BBCJPW93}, quantum
computation \cite{RB01}, and violations of Bell inequalities
\cite{Bell64, ADR82, TBZG98, WJSWZ98,RMSIMW01, GPKBZAZ07, MMMOM08,
BBGKLLS08}. However, Bohr argued that similar paradoxical examples
occur every time we compare different experimental arrangements,
without the need of entanglement nor composite systems
\cite{Bohr35}. The Kochen-Specker (KS) theorem \cite{Specker60,
Bell66, KS67} illustrates Bohr's intuition with great precision. The
KS theorem states that, for every physical system there is always a
finite set of tests such that it is impossible to assign them
predefined noncontextual results in agreement with the predictions
of quantum mechanics \cite{Specker60, KS67}. Remarkably, the proof
of the KS theorem \cite{KS67} requires neither a composite system
nor any special quantum state: it holds for any physical system with
more than two internal levels (otherwise the notion of
noncontextuality becomes trivial), independent of its state. It has
been discussed for a long time whether or not the KS theorem can be
translated into experiments \cite{CG98, Meyer99}. Recently, however,
quantum contextuality has been tested with single photons
\cite{SZWZ00, HLZPG03} and single neutrons \cite{HLBBR03} in
specific states.

Very recently it has been shown that the KS theorem can be converted
into experimentally testable state-dependent \cite{CFRH08} and
state-independent \cite{Cabello08} violations of inequalities
involving correlations between compatible measurements. For single
systems, only a state-dependent violation for a specific state of
single neutrons has been reported \cite{BKSSCRH09}. A
state-independent violation has been observed only in composite
systems of two $^{40}$Ca$^+$ trapped ions \cite{KZGKGCBR09}.
Following the spirit of the original KS theorem, which deals with
the problem of hidden variables in single systems, we report the
first state-independent violation for single-particle systems.

Any theory in which the nine observables $A, B, C, a, b, c, \alpha,
\beta$, and $\gamma$ have predefined noncontextual outcomes $-1$ or
$+1$, must satisfy the following inequality \cite{Cabello08}:
\begin{equation}
\chi \equiv \langle A B C \rangle + \langle a b c \rangle + \langle
\alpha \beta \gamma \rangle +\langle A a \alpha \rangle + \langle B
b \beta \rangle - \langle C c \gamma \rangle \le 4, \label{second}
\end{equation}
where $\langle A B C \rangle$ denotes the ensemble average of the
product of the three outcomes of measuring the mutually compatible
observables $A$, $B$, and $C$. Surprisingly, for any
four-dimensional system, there is a set of observables for which the
prediction of quantum mechanics is $\chi=6$ for any quantum state of
the system \cite{Cabello08}. The purpose of this experiment is to
test this prediction on different quantum states of a
single-particle system.

A physical system particularly well suited for this purpose is the
one comprising a single photon carrying two qubits of quantum
information: the first qubit is encoded in the spatial path $s$ of
the photon, and the second qubit in the polarization $p$. The
quantum states $|0\rangle_s=|t\rangle_s$ and
$|1\rangle_s=|r\rangle_s$, where $t$ and $r$ denote the transmitted
and reflected paths of the photon, respectively, provide a basis for
describing any quantum state of the photon's spatial path.
Similarly, $|0\rangle_p=|H\rangle_p$ and $|1\rangle_p=|V\rangle_p$,
where $H$ and $V$ denote horizontal and vertical polarization,
respectively, provide a basis for describing any quantum state of
the photon's polarization.

A suitable choice of observables giving $\chi = 6$ is the following
\cite{Cabello08}:
\begin{eqnarray}
&&\;\;A=\sigma_z^s,\;\;\;\;\;\;\;\;\;\;\;\;\;\;
 B=\sigma_z^p,\;\;\;\;\;\;\;\;\;\;\;\;
 C=\sigma_z^s \otimes \sigma_z^p, \nonumber \\
&&\;\;a=\sigma_x^p,\;\;\;\;\;\;\;\;\;\;\;\;\;\;\;\;
 b=\sigma_x^s,\;\;\;\;\;\;\;\;\;\;\;\;
 c=\sigma_x^s \otimes \sigma_x^p, \nonumber \\
&&\alpha=\sigma_z^s \otimes \sigma_x^p,\;\;\;\;\;\;\;
 \beta=\sigma_x^s \otimes \sigma_z^p,\;\;\;\;\;\;
 \gamma=\sigma_y^s \otimes \sigma_y^p,
\label{observables}
\end{eqnarray}
where $\sigma_z^s$ denotes the Pauli matrix along the $z$ direction
of the spatial path qubit, $\sigma_x^p$ denotes the Pauli matrix
along the $x$ direction of the polarization qubit, and $\otimes$
denotes tensor product.


\begin{figure}[tb]
\center
\includegraphics[width=0.70\linewidth]{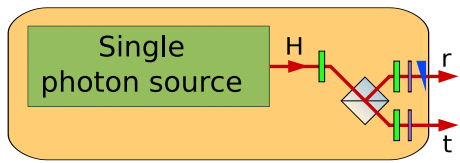}
\caption{Preparation of the polarization-spatial path encoded states
of single photons. The setup consists of a source of $H$-polarized
single photons followed by a half wave plate (HWP) and a polarizing
beam splitter (PBS), allowing any probability distribution of a
photon in the paths $t$ and $r$. The wedge (W) placed in one of the
paths adds an arbitrary phase shift between both paths. A HWP and a
quarter wave plate (QWP) in each path allow us to rotate the outputs
of the PBS to any polarization. Symbol definitions are given at the
bottom of Fig. \ref{Blocks}.} \label{Preparation}
\end{figure}



\begin{figure}[tb]
\center
\includegraphics[width=1.00\linewidth]{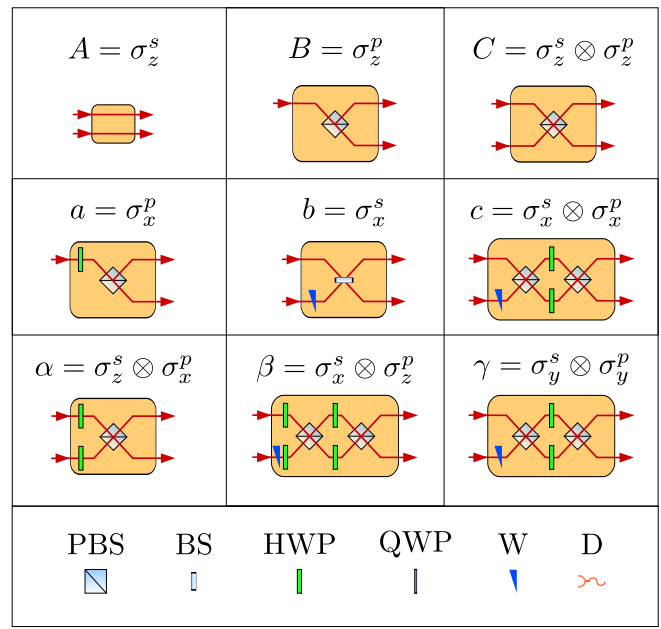}
\caption{Devices for measuring the nine observables
(\ref{observables}). A measurement of $A$ requires only to
distinguish between paths $t$ and $r$. For measuring $b$, note that
its eigenstates are $(|t\rangle \pm |r\rangle)/ \sqrt{2}$ and they
need to be mapped to the paths $t$ and $r$; this is accomplished by
interference with the help of an additional $50/50$ beam splitter
(BS) and a wedge. The measurements of $a$ and $B$ are standard
polarization measurements using a PBS and a HWP. Observables $C$,
$c$, $\alpha$, $\beta$, and $\gamma$ are the product of a spatial
path and a polarization observable $\sigma_i^s \otimes \sigma_j^p$.
Each of these observables has a four-dimensional eigenspace, but
since the observables need to be rowwise and columnwise compatible,
only their common eigenstates can be used for distinguishing the
eigenvalues. This implies that $C$, $c$, and $\gamma$ can be
implemented as Bell measurements with different distributions of the
Bell states. Similarly, $\alpha$ and $\beta$ are Bell measurements
preceded by a polarization rotation. In this way $\gamma$ is
compatible also with $\alpha$ and $\beta$.} \label{Blocks}
\end{figure}


To generate polarization-spatial path encoded single-photon states,
we used the setup described in Fig.~\ref{Preparation}. We
experimentally tested the value of $\chi$ for 20 different quantum
states. It is of utmost importance for the experiment that the
measurements of each of the nine observables in (\ref{observables})
are context independent \cite{Cabello08}, in the sense that the
measurement device used for the measurement of, e.g., $B$ must be
the same when $B$ is measured with the compatible observables $A$
and $C$, and when $B$ is measured with $b$ and $\beta$, which are
compatible with $B$ but not with $A$ and $C$. For the experiment we
used the measurement devices described in Fig.~\ref{Blocks}, which
satisfy this requirement.



\begin{figure}[tb]
\center
\includegraphics[width=1.00\linewidth]{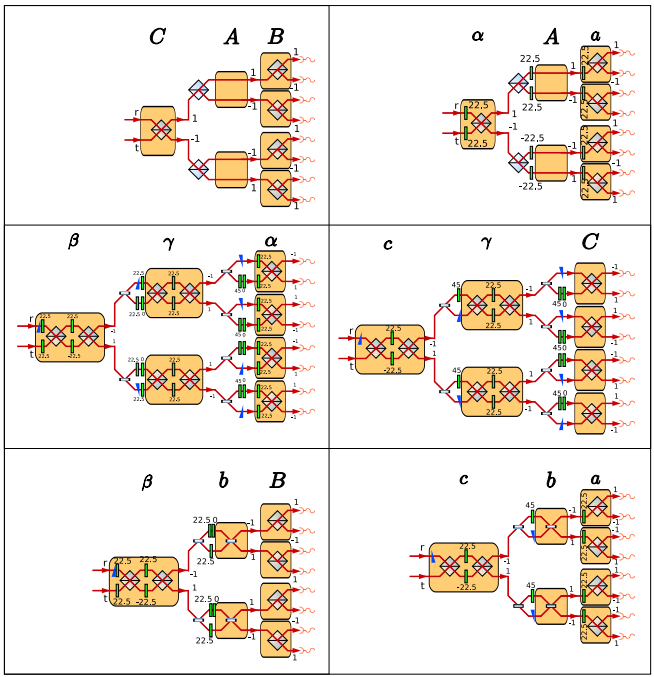}
\caption{Setups for measuring the six sets of observables to test
inequality (\ref{second}). We explicitly describe the setup for
measuring $C$, $A$, and $B$; the description of the other setups is
obtained by replacing $C$, $A$, and $B$ with the corresponding
observables. The seven boxes are single-observable measuring devices
(see Fig.~\ref{Blocks}). The photon, prepared in a specific state,
enters the device for measuring $C$ through the device's input and
follows one of the two possible outcomes. A detection of the photon
in one of these outputs would make the measurement of the next
observable impossible. Instead, we placed, after each of the two
outputs of the $C$-measuring device, a device for measuring the
second observable, $A$ (we thus used two identical $A$-measuring
devices). Similarly, we also placed, after each of the four outputs
of the $A$-measuring devices, a device for measuring the third
observable, $B$ (we thus used four identical $B$-measuring devices).
Note that we need to recreate the eigenstates of the measured
observable before entering the next observable, since our
single-observable measuring devices map eigenstates to a fixed
spatial path and polarization. Finally, we placed a single-photon
detector (D) after each of the eight outputs of the four
$B$-measuring devices. An individual photon passing through the
whole arrangement is detected only by one of the eight detectors,
which indicates which one of the eight combinations of results for
$C$, $A$, and $B$ is obtained.}\label{configurations}
\end{figure}



\begin{figure}[t]
\center
\includegraphics[width=1.00\linewidth]{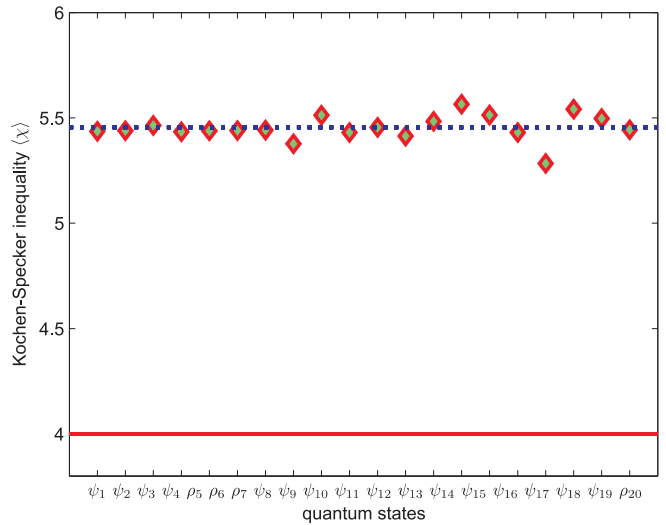}
\caption{State independence of the violation of the inequality $\chi
 \le 4$. The value of $\chi$ was tested for 20 different quantum
 states:
 four pure states with maximum internal entanglement between
 the spatial path and polarization which would maximally
 violate a Clauser-Horne-Shimony-Holt-Bell-like inequality \cite{CHSH69}
 (states $|\psi_1\rangle$--$|\psi_4\rangle$),
 one mixed state with partial internal entanglement which would violate a Clauser-Horne-Shimony-Holt-Bell-like inequality ($\rho_{5}$),
 one mixed state with partial internal entanglement which would not violate a Clauser-Horne-Shimony-Holt-Bell-like inequality ($\rho_{6}$),
 one mixed state without internal entanglement according to the Peres-Horodecki criterion \cite{Peres96, HHH96} ($\rho_{7}$),
 12 pure states without internal entanglement ($|\psi_{8}\rangle$--$|\psi_{19}\rangle$),
 and a maximally mixed state ($\rho_{20}$).
 The explicit expression of each state is given in Table \ref{Table}.
 The red solid line indicates the classical upper bound.
 The blue dashed line at $5.4550$ indicates the average value of $\chi$ over all the 16 pure states.
 }\label{Violation}
\end{figure}



\begin{figure}[tb]
\center
\includegraphics[width=1.00\linewidth]{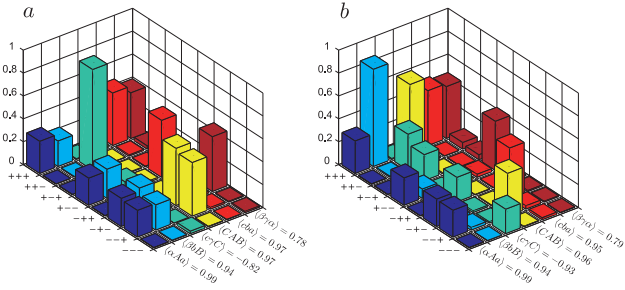}
\caption{Correlation measurements of all terms in the inequality
(\ref{second}) for the states $|\psi_{3}\rangle$ ($a$) and
$|\psi_{14}\rangle$ ($b$). The figures show experimentally estimated
probabilities for detecting a photon in each of the eight detectors.
A photon detection corresponds to certain values ($\pm 1$) for the
three measured dichotomic observables. For example, the bar height
at ($+++;\alpha A a$) represents the probability to obtain the
results $\alpha, A, a=+1$, and similarly ($++-;\alpha A a$)
represents $\alpha, A=+1$ and $a=-1$. The expectation values for
each measurement are also given.}\label{States}
\end{figure}



\begin{table}[b]
\begin{center}
\caption{Experimental values of $\langle CAB \rangle + \langle cba
\rangle + \langle \beta \gamma \alpha \rangle + \langle \alpha Aa
\rangle + \langle \beta b B \rangle - \langle c \gamma C \rangle$
for 20 quantum states. The average value is $5.4550\pm0.0006$ and on
average we violate the inequality with $655$ standard deviations
(SDs).} \label{Table}
\begin{ruledtabular}
\begin{tabular}{c c r}
State & Expectation value & SD \\ \hline
$\ket{\psi_1} = \frac{1}{\sqrt{2}}(\ket{t}\ket{H}+\ket{r}\ket{V})$                                                   & $5.4366$ $\pm$ $0.0012$ & $1169$ \\
$\ket{\psi_2} = \frac{1}{\sqrt{2}}(\ket{t}\ket{H}-\ket{r}\ket{V})$                                                   & $5.4393$ $\pm$ $0.0023$ & $621 $ \\
$\ket{\psi_3} = \frac{1}{\sqrt{2}}(\ket{t}\ket{V}+\ket{r}\ket{H})$                                                   & $5.4644$ $\pm$ $0.0029$ & $498 $ \\
$\ket{\psi_4} = \frac{1}{\sqrt{2}}(\ket{t}\ket{V}-\ket{r}\ket{H})$                                                   & $5.4343$ $\pm$ $0.0026$ & $561 $ \\
$\rho_{5} = \frac{13}{16} |\psi_1\rangle \langle\psi_1| + \frac{1}{16} \sum_{j=2}^{4} |\psi_j\rangle \langle\psi_j|$ & $5.4384$ $\pm$ $0.0010$ & $1386$ \\
$\rho_{6} = \frac{5}{8} |\psi_1\rangle \langle\psi_1| + \frac{1}{8} \sum_{j=2}^{4} |\psi_j\rangle \langle\psi_j|$    & $5.4401$ $\pm$ $0.0010$ & $1509$ \\
$\rho_{7} = \frac{7}{16} |\psi_1\rangle \langle\psi_1| + \frac{3}{16} \sum_{j=2}^{4} |\psi_j\rangle \langle\psi_j|$  & $5.4419$ $\pm$ $0.0010$ & $1433$ \\
$\ket{\psi_{8}} = \ket{t}\ket{H}$                                                                                    & $5.3774$ $\pm$ $0.0020$ & $676 $ \\
$\ket{\psi_{9}} = \ket{t}\ket{V}$                                                                                    & $5.5131$ $\pm$ $0.0032$ & $475 $ \\
$\ket{\psi_{10}} = \ket{r}\ket{H}$                                                                                   & $5.4306$ $\pm$ $0.0031$ & $465 $ \\
$\ket{\psi_{11}} = \ket{r}\ket{V}$                                                                                   & $5.4554$ $\pm$ $0.0017$ & $850 $ \\
$\ket{\psi_{12}} = \frac{1}{\sqrt{2}}\ket{t}(\ket{H}+\ket{V})$                                                       & $5.4139$ $\pm$ $0.0015$ & $960 $ \\
$\ket{\psi_{13}} = \frac{1}{\sqrt{2}}\ket{t}(\ket{H}+i\ket{V})$                                                      & $5.4835$ $\pm$ $0.0022$ & $667 $ \\
$\ket{\psi_{14}} = \frac{1}{\sqrt{2}}(\ket{t}+\ket{r})\ket{H}$                                                       & $5.5652$ $\pm$ $0.0032$ & $489 $ \\
$\ket{\psi_{15}} = \frac{1}{\sqrt{2}}(\ket{t}+i\ket{r})\ket{H}$                                                      & $5.5137$ $\pm$ $0.0036$ & $419 $ \\
$\ket{\psi_{16}} = \frac{1}{2}(\ket{t}+\ket{r})(\ket{H}+\ket{V})$                                                    & $5.4304$ $\pm$ $0.0014$ & $1029$ \\
$\ket{\psi_{17}} = \frac{1}{2}(\ket{t}+i\ket{r})(\ket{H}+\ket{V})$                                                   & $5.2834$ $\pm$ $0.0019$ & $674 $ \\
$\ket{\psi_{18}} = \frac{1}{2}(\ket{t}+\ket{r})(\ket{H}+i\ket{V})$                                                   & $5.5412$ $\pm$ $0.0032$ & $475 $ \\
$\ket{\psi_{19}} = \frac{1}{2}(\ket{t}+i\ket{r})(\ket{H}+i\ket{V})$                                                  & $5.4968$ $\pm$ $0.0032$ & $462 $ \\
$\rho_{20} = \frac{1}{4} \sum_{j=1}^{4} |\psi_j\rangle\langle\psi_j|$                                                & $5.4437$ $\pm$ $0.0012$ & $1229$ \\
\end{tabular}
\end{ruledtabular}
\end{center}
\end{table}


For a sequential measurement of three compatible observables on the
same photon, we used the single-observable measuring devices in
Fig.~\ref{Blocks}, appropriately arranged as described in
Fig.~\ref{configurations}. Since the predictions of both
noncontextual hidden variable theories and quantum mechanics do not
depend on the order of the compatible measurements, we chose the
most convenient order for each set of observables (e.g., we measured
$CBA$ instead of $ABC$). This was usually the configuration which
minimized the number of required interferometers and hence maximized
the visibility. Specifically, we measured the averages $\langle CAB
\rangle$, $\langle c b a \rangle$, $\langle \beta \gamma \alpha
\rangle$, $\langle \alpha A a \rangle$, $\langle \beta b B \rangle$,
and $\langle c \gamma C \rangle$, as described in
Fig.~\ref{configurations}.


Our single-photon source was an attenuated stabilized narrow
bandwidth diode laser emitting at 780 nm and offering a long
coherence length. The laser was attenuated so that the two-photon
coincidences were negligible. The mean photon number per time window
was $0.058$.

All the interferometers in the experimental setup are based on free
space displaced Sagnac interferometers, which possess a very high
stability. We have reached a visibility above $99\%$ for phase
insensitive interferometers, and a visibility ranging between $90\%$
and $95\%$ for phase sensitive interferometers.

Our single-photon detectors were Silicon avalanche photodiodes
calibrated to have the same detection efficiency. All single counts
were registered using an eight-channel coincidence logic with a time
window of $1.7$ ns.


To test the prediction of a state-independent violation, we repeated
the experiment on 20 quantum states of different purity and
entanglement. For each pure state, we checked each of the six
correlations in inequality (\ref{second}) for about $1.7 \times
10^7$ photons. The results for the mixed states were obtained by
suitably combining pure state data. Fig.~\ref{Violation} shows that
a state-independent violation of inequality $\chi \le 4$ occurs,
with an average value for $\chi$ of $5.4521$. Because of
experimental imperfections, the experimental violation of the
inequality falls short of the quantum-mechanical prediction for an
ideal experiment ($\chi = 6$).


The main systematic error source was due to the large number of
optical interferometers involved in the measurements, the nonperfect
overlapping of the light modes and the polarization components. The
errors were deduced from propagated Poissonian counting statistics
of the raw detection events. The number of detected photons was
about $1.7 \times 10^6$ per second. The measurement time for each of
the six sets of observables was $10$ s for each state.


In Fig.~\ref{States} we also present measurement results for each
experimental setup for the maximally entangled state
$|\psi_3\rangle$ and the product state $|\psi_{14}\rangle$ defined
in Table \ref{Table}. Probabilities for each outcome as well as
values of the correlations are shown. The overall detection
efficiency of the experiment, defined as the ratio of detected to
prepared photons, was $\eta = 0.50$. This value was obtained
considering that the detection efficiency of the single-photon
detectors is $55\%$ and the fiber coupling is $90\%$. Therefore, the
fair sampling assumption (i.e., the assumption that detected photons
are an unbiased subensemble of the prepared photons) is needed to
conclude a violation of the inequality. This is the same assumption
as is adopted in all previous state-dependent experimental
violations of classical inequalities with photons \cite{ADR82,
TBZG98, WJSWZ98, GPKBZAZ07, MMMOM08, SZWZ00, HLZPG03} and neutrons
\cite{HLBBR03, BKSSCRH09}.


In conclusion, our results show that experimentally observed
outcomes of measurements on single photons cannot be described
by noncontextual models. A remarkable feature of this
experiment is that the quantum violation of a classical
inequality requires neither entangled states nor composite
systems. It occurs even for single systems which cannot have
entanglement. Further on, it occurs for any quantum state, even
for maximally mixed states, like $\rho_{20}$ in Fig.
\ref{Violation}, which are usually considered ``classical''
states. This shows that entanglement is not the only essence of
quantum mechanics which distinguishes the theory from classical
physics; consequently, entanglement might not be the only
resource for quantum information processing. Quantum
contextuality of single quantum systems submitted to a sequence
of compatible measurements might be an equally powerful,
simpler and more fundamental resource.


\begin{acknowledgments}
We thank Y. Hasegawa and J.-\AA. Larsson for comments, and
acknowledge support by the Swedish Research Council (VR), the
Spanish MCI Project No.\ FIS2008-05596, and the Junta de
Andaluc\'{\i}a Excellence Project No.\ P06-FQM-02243.
\end{acknowledgments}



\end{document}